\documentclass[aps,pra,superscriptaddress,amssymb,showpacs]{revtex4}
\usepackage{graphicx}
\usepackage{bm}
\usepackage{amsmath}

\begin{document}

\title{Statistical Properties of Ideal Ensemble of Disordered 1$D$ Steric Spin-Chains }

\author{A.~S.~Gevorkyan}
\email[]{g_ashot@sci.am}
\affiliation{Institute for Informatics and Automation Problems, NAS
of Armenia}

\author{H.~G.~ Abajyan}
\email[]{habajyan@ipia.sci.am} \affiliation{Institute for
Informatics and Automation Problems, NAS of Armenia}

\author{H.~S.~Sukiasyan}
\email[]{haikarin@netsys.am}
\affiliation{Institute of Mathematics,  NAS of Armenia }

\begin{abstract}
The statistical properties of ensemble of disordered 1$D$  steric
spin-chains (SSC)  of various length are investigated. Using 1$D$
spin-glass type classical Hamiltonian, the recurrent trigonometrical
equations  for stationary points and corresponding conditions for
the construction of stable 1$D$ SSCs are found. The ideal ensemble
of spin-chains  is analyzed and the latent interconnections between
random angles and interaction constants for each set of three
nearest-neighboring spins are found.  It is analytically proved and
by numerical calculation is shown that the interaction constant
satisfies Le\'{v}y's alpha-stable distribution law. Energy
distribution in ensemble is calculated depending on different
conditions of possible polarization of spin-chains. It is
specifically shown that the dimensional effects in the form of set
of local maximums  in the energy distribution arise when the number
of spin-chains $M<<N^2_x$ (where $N_x$ is number of spins in a
chain) while in the case when $M\propto N^2_x$  energy distribution
has one global maximum and ensemble of spin-chains satisfies
Birkhoff's ergodic theorem. Effective algorithm for parallel
simulation of problem which includes calculation of different
statistic parameters of 1$D$ SSCs ensemble is elaborated.
\end{abstract}



\maketitle

\section{Formulation of the problem}
Let us consider classical ensemble of disordered  1$D$  steric
spin-chains (SSC), where  it is supposed that interactions between
spin-chains are absent (later it will be called an ideal ensemble)
and that there are $N_x$ spins in an each chain.  Despite some
ideality of the model it can be interesting enough and rather
convenient for investigation of a number of important  and difficult
applied problems of physics, chemistry, material science, biology,
evolution, organization dynamics, hard-optimization, environmental
and social structures, human logic systems, financial mathematics
etc (see for example \cite{Young,Bov,Fisch,Tu,Chary,Baake}). As was
shown by authors spin-glass model can be used for investigation of
media's properties on scales of space-time periods of an external
fields at conditions far from   a usual equilibrium of media
\cite{gev}.

Mathematically  mentioned type of ideal ensemble can be generated by
1$D$ Heisenberg spin-glass Hamiltonian without external field
\cite{Bind,Mezard,Young}:

\begin{figure}
\center
\includegraphics[height=90mm,width=120mm]{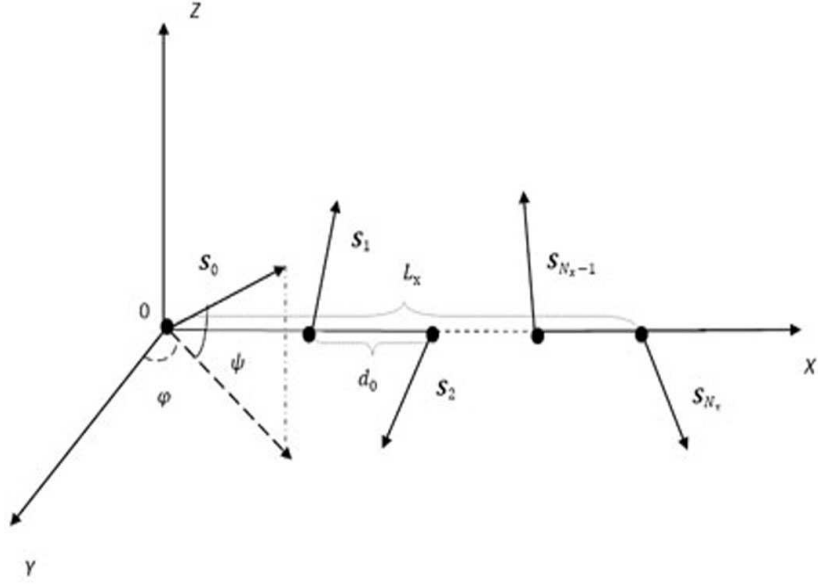}
 \vspace{-0.5 cm} \caption{A stable  $1D$ steric spin-chain with random interactions and the length of $L_x$.
The spherical angles $\varphi$ and $\psi$  describe the spatial
orientation of $\textbf{\emph{S}}_0$ spin, the pair of angles
$(\varphi_i, \psi_i)$ correspondingly defines the spatial
orientation of the spin $\textbf{\emph{S}}_i$, the distance between
two neighboring spins in $1D$ lattice is $d_0$.} \label{fig1}
\end{figure}

\begin{eqnarray}
H_0(N_x)=-\sum_{i=0}^{N_x-1}
J_{i\,i+1}{\textbf{\emph{S}}}_{i}{\textbf{\emph{S}}}_{i+1}.
\label{01}
\end{eqnarray}
where ${\textbf{\emph{S}}}_{i}$  describes the $i$-th spin which
is a unit length vector and has a random orientation.  In the
expression (\ref{01}) $J_{i\,i+1}$ characterizes a random
interaction constant between  $i$ and $i+1$ spins, which can have
positive and negative values as well \cite{EdwAnd}.

In other words we consider the  mathematical model of spin-chains
ensemble where every spin-chain is like a regular 1$D$ lattice with
the length $L_x=d_0N_x$, where spins are put on nodes of  lattice
and interactions between them are random (see FIG 1).

The distribution of  spin-spin interaction constant $W(J)$ is
chosen from  considerations of convenience and as a rule  it is a
Gauss-Edwards-Anderson model \cite{EdwAnd} (see also \cite{Bind}):
\begin{eqnarray}
W(J)=\frac{1}{\sqrt{2\pi(\Delta{J})^2}}\exp
\biggl\{-\frac{\bigl(J-J_0\bigr)^2} {2(\Delta{J})^2}\biggr\},
 \label{02}
\end{eqnarray}
where $J_0=\bigl<J\bigr>_{av}$ and
$\bigl(\Delta{J}\bigr)^2=\bigl<J^2\bigr>_{av}
-\bigl<J\bigr>_{av}^2$.

Let us recall that $J_{0}$ and $\Delta {J}$ for this model are
independent from the distance and scaled with the spin number
$N_{x}$ as:
\begin{eqnarray}
\bigl<J\bigr>_{av}=J_0\propto{N_x^{-1}},\qquad
\Delta{J}\propto{N_x^{-1/2}},
 \label{03}
\end{eqnarray}
in order to ensure a sensible thermodynamic limit.
$\bigl<...\bigr>_{av}$ in Eqs. (\ref{02}) and (\ref{03}) describes
the averaging procedure. Below we will investigate the issue of how
much lawful the choice of this model is.

For further investigations it is useful to rewrite the Hamiltonian
(\ref{01}) in spherical coordinates (see FIG 1):
\begin{eqnarray}
H_0(N_x)=-\sum_{i=0}^{N_x-1}
J_{i\,i+1}\bigl[\cos\psi_i\cos\psi_{i+1}\cos(\varphi_i-\varphi_{i+1})
+\sin\psi_i\sin\psi_{i+1}\bigr].
 \label{04}
\end{eqnarray}
A stationary point of the Hamiltonian is given by the system of
trigonometrical  equations:
\begin{equation}
\frac{\partial{H_0}}{\,\,\partial\psi_i}=0,\qquad\qquad
\frac{\partial{H_0}}{\,\,\partial\varphi_i}=0,
 \label{05}
\end{equation}
where ${\Theta }_{i}=(\psi _{i},\varphi _{i})$ are  angles of $i$-th
spin  in the spherical coordinates system ($\psi _{i}$ is a polar
and $\varphi _{i}$ is an azimuthal angles), $\mathbf{\Theta }=({
\Theta _{1}},{\Theta _{2}}....{\Theta _{N_{x}}})$ respectively
describe the angular part of a spin-chain configuration.

Now using expression (\ref{04})  and equations (\ref{05}) it is easy
to find the following system of  trigonometrical equations:
\begin{eqnarray}
\sum_{\nu=i-1;\,\,\nu\neq i}^{i+1} J_{\nu
i}\bigl[\sin\psi_{\nu}-\tan\psi_i\cos\psi_{\nu}\cos(\varphi_i-\varphi_{\nu})
\bigr] =0,
\nonumber\\
\sum_{\nu=i-1;\,\,\nu\neq i}^{i+1}J_{\nu
i}\,\cos\psi_{\nu}\sin(\varphi_i-\varphi_{\nu})=0, \,\,\,\qquad\,
J_{\nu i}\equiv J_{i\nu} .  \label{06}
\end{eqnarray}

 In case when  all the interaction constants between
$i$-th spin with  its nearest-neighboring spins
$J_{i-1\,i}$,\thinspace \thinspace\ $J_{i\,i+1}$ and angle
configurations $\bigl(\psi _{i-1},\varphi _{i-1}\bigr)$,\thinspace
\thinspace $\bigl(\psi _{i},\varphi _{i}\bigr)$ are known,  it is
possible to explicitly calculate the pair of angles ${\Theta
_{i+1}}=\bigl(\psi _{i+1},\varphi_{i+1}\bigr)$. Correspondingly, the
$i$-th spin will be in the ground state (in the state of minimum
energy) if in the stationary point ${\Theta _{i}^{0}}= \bigl(\psi
_{i}^{0},\varphi _{i}^{0}\bigr)$ the following conditions are
satisfied:
\begin{equation}
A_{\psi_i\psi_i}({\Theta_i^0})>0,\qquad
A_{\psi_i\psi_i}({\Theta_i^0})\,A_{\varphi_i\varphi_i}({\Theta_i^0})
-A_{\psi_i\phi_i}^2({\Theta_i^0})>0,
 \label{07}
\end{equation}
where $A_{\alpha_i\alpha_i}({\Theta_i^0})
={\partial^2{H_0}}/{\partial\alpha_i^2}, \quad
A_{\alpha_i\beta_i}({\Theta_i^0})=A_{\beta_i \alpha_i}({\Theta_i^0
})= {\partial^2{H_0}}/{\partial\alpha_i\partial\beta_i}$, in
addition:
$$
A_{\psi_i\psi_i}({\Theta_i^0})\,
=\,\biggl\{\,\sum_{\nu=i-1;\,\,\nu\neq i}^{i+1}J_{\nu
i}\bigl[\cos\psi_{\nu}\cos(\varphi_{\nu}-\varphi_i^0)
+\tan\psi_i^0\sin\psi_{\nu}\bigr]\biggr\}\cos\psi_i^0,\,\,\,\,
$$
\begin{eqnarray}
A_{\varphi_i\varphi_i}({\Theta_i^0})=\biggl\{\,\sum_{\nu=i-1;\,\,\nu\neq
i}^{i+1} J_{\nu
i}\cos\psi_{\nu}\cos(\varphi_{\nu}-\varphi_i^0)\biggr\}\cos\psi_i^0,
\qquad\qquad\qquad
\nonumber\\
 A_{\psi_i\phi_i}({\Theta_i^0})=\biggl\{\,\sum_{\nu=i-1;\,\,\nu\neq
i}^{i+1} J_{\nu
i}\cos\psi_{\nu}\sin(\varphi_{\nu}-\varphi_i^0)\,\biggr\}\sin\psi_i^0.
\qquad\qquad\qquad
\label{08}
\end{eqnarray}

Taking  into account the second equation in (\ref{06}) we can reduce
condition (\ref{07})  to the following kind:
\begin{equation}
A_{\psi_i\psi_i}({\Theta_i^0})>0,\qquad\qquad
A_{\varphi_i\varphi_i}({\Theta_i^0})>0.
 \label{09}
\end{equation}

So, with the help of  Eq.s   (\ref{06}) and conditions (\ref{09})
huge  number of stable $1D$ SSCs may be  calculated  and on its
basis it is possible to further construct the statistical properties
of 1$D$ SSCs ensemble. It is important to note that the average
polarization of 1$D$ SSCs ensemble  is supposed to be equal to zero.

Now we can  construct  the distribution function of energy  in 1$D$
SCCs ensemble. To this effect it is useful to divide the
nondimensional energy axis $\varepsilon=\epsilon/\delta\epsilon$
into regions $0>\varepsilon _{0}>...>\varepsilon _{n}$, where $n>>1$
and $\epsilon$ is a real energy axis. The number of stable 1$D$ SSC
configurations with length of $L_{x}$ in the range of energy $
[\varepsilon -\delta \varepsilon,\varepsilon +\delta \varepsilon ]$
will be denoted by $M_{L_{x}}(\varepsilon )$ while the number of all
stable 1$D$ SSC configurations - correspondingly by symbol
$M_{L_{x}}^{full}= \sum_{j=1}^{n}M_{L_{x}}(\varepsilon _{j})$.
Accordingly, the energy distribution function into the 1$D$ SSCs
ensemble may be defined by expressions:
\begin{equation}
F_{L_x}(\varepsilon;d_0(T))=M_{L_x}(\varepsilon)/M_{L_x}^{full},
\label{10}
\end{equation}
where distribution function is normalized to unit:
$$
 \lim_{n\to\infty}\sum^n_{j=1}
F_{L_x}(\varepsilon_j;d_0(T))\delta \varepsilon_j=
\int^{\,0}_{-\infty}F_{L_x}(\varepsilon;d_0(T))d\varepsilon=1.
$$
 By similar way we  can define also   distributions for
polarization and for a spin-spin interaction constant.

\section{Algorithm of 1$D$ SSCs Ideal  Ensemble Simulation  }
Now our aim is elaboration of algorithm for parallel
simulation of ideal ensemble of $1D$ SSCs.\\
Using equations (\ref{06}) for stationary points of Hamiltonian
$H_0(N_x)$ we can find the following equations system:
\begin{eqnarray}
J_{i-1\,
i}\bigl[\sin\psi_{i-1}-\tan\psi_i\cos\psi_{i-1}\cos(\varphi_i-\varphi_{i-1})
\bigr]+J_{i\, i+1}\bigl[\sin\psi_{i+1}&
\nonumber\\
-\tan\psi_i\cos\psi_{i+1}\cos(\varphi_i-\varphi_{i+1}) \bigr]=0,&
\nonumber\\
J_{i-1\, i}\,\cos\psi_{i-1}\sin(\varphi_i-\varphi_{i-1})\,+J_{i+1\,
i}\,\cos\psi_{i+1}\sin(\varphi_i-\varphi_{i+1})=0.&
 \label{11}
\end{eqnarray}
After designations:
\begin{equation}
x =\cos\psi_{i+1},\qquad y=\sin(\varphi_i-\varphi_{i+1}),
 \label{12}
\end{equation}
the system (\ref{11}) may be transformed to the following form:
\begin{eqnarray}
 C_1+J_{i\,
i+1}\bigl[\sqrt{1-x^2}-\tan\psi_i\,x \sqrt{1-y^2} \bigr]=0, \qquad
C_2+J_{i\, i+1}\,x\,y=0,
 \label{13}
\end{eqnarray}
where parameters $C_1$ and $C_2$ are defined by expressions:
\begin{eqnarray}
C_1=J_{i-1\,i}\bigl[\sin\psi_{i-1}-\tan\psi_i\cos\psi_{i-1}\cos(\varphi_i-\varphi_{i-1})
\bigr],\nonumber\\
 C_2=J_{i-1\,i}\cos\psi_{i-1}
\sin(\varphi_i-\varphi_{i-1}).\qquad\qquad\qquad\qquad\,\,
 \label{14}
\end{eqnarray}
From the system of equations (\ref{13}) we can find the equation for
the unknown variable $y$:
\begin{equation}
C_1y+C_2\sqrt{1-y^2}\tan\psi_i+\sqrt{J_{i\, i+1}^2y^2-C_2^2}=0.
 \label{15}
\end{equation}
We can transform the  equation (\ref{15}) to the following equation
of fourth order:
\begin{equation}
\label{16} \bigl[A^2+4C_1^2C_2^2\sin\psi_i\bigl]
y^4-2\bigl[AC_2^2+2C_1C_2^2\sin^2\psi_i\bigr]y^2+C_2^4=0,
\end{equation}
where
\begin{equation}
\label{17} A=J^2_{i\,i+1}\cos^2\psi_i-C_1^2+C_2^2\sin^2\psi_i.
\end{equation}
Discriminant of equation (\ref{16}) is equal to:
$$
D=C_2^4\bigl(A+2C_1\sin^2\psi_i\bigr)^2-C_2^4\bigl(A^2+4C_1^2C_2^2\sin^2\psi_i\bigr)
$$
$$
= 4C^4_2C^2_1 \sin^2\psi_i\bigl(A + C^2_1 \sin^2\psi_i -
C^2_2).\qquad\qquad
$$
From the condition of nonnegativity of discriminant $D\geq 0$ we can
find the following condition:
\begin{equation}
A + C^2_1 \sin^2\psi_i - C^2_2\geq 0.
 \label{18}
\end{equation}
Further substituting  the value of  $A$ from (\ref{17}) into
(\ref{18}) we can find the new condition to which  the interaction
constant between two successive spins should satisfy:
\begin{equation}
J_{i\,i+1}^2 \geq C^2_1 + C^2_2.
 \label{19}
\end{equation}
Now we can write the following expressions for unknown variables $x$
and $y$:
\begin{align}
 x^2 & =\frac{C_2^2}{J_{i\,i+1}^2y^2}, \nonumber\\
  y^2 & =
C_2^2\,\frac{\cos^2\psi_iJ_{i\,i+1}^2\pm
2C_1\sin\psi_i\cos\psi_i\sqrt{J_{i\,i+1}^2- C_1^2-C_2^2}+
C_3+2C_1^2\sin^2\psi_i}{\cos^4\psi_iJ_{i\,i+1}^4+
2C_3\cos^2\psi_iJ_{i\,i+1}^2+(C_1^2+\sin^2\psi_iC_2^2)^2},\label{20}
\end{align}
where $C_3=-C_1^2+C_2^2\,\sin^2\psi_i.$\\
 Finally  taking into account
designations (\ref{12}) we can find  new  conditions of restriction
of  the calculated angles   $\bigl(\varphi_{i+1},\psi_{i+1}\bigr)$:
\begin{equation}
0\leq x^2\leq 1, \qquad 0\leq y^2\leq 1.
 \label{21}
\end{equation}
These conditions are very important for elaborating correct and
effective algorithm for numerical simulations.

\subsection{Algorithm description}
This is parallel algorithm for simulation of 1$D$ SSCs ensemble,
which consists of separate iterative calculations of nodes in 1$D$
SSC. The first and second nodes are initialized randomly, then
$i$-th node is obtained from $(i-2)$-th and $(i-1)$-th layers nodes.
Every node
contains the following information:\\
$\varphi$-polar angle,\\
$\psi$-azimuthal angle,\\
$J$-interaction coefficient,\\

\vspace{0.05 cm}

The following parameters are initializes in the following way:\\
$\varphi_0$ and $\varphi_1$ - \textbf{rand()}\textbf{$^\ast2^\ast\pi^\ast R$};\\
$\psi_0$ and $\psi_1$ - \textbf{acos (rand())};\\
$J_{0\,1}$ - \textbf{rand()};\\

where \textbf{rand()} function generates uniformly distributed
random numbers on the interval $(0,1)$.\\

The algorithm pseudo-code is following:
 \vspace{0.25 cm}

\quad{//} generate $n$ separate independent sets of problem in
parallel

\qquad for $i=1:N_x$

\qquad\qquad   for $j=1:R$  \quad{//} regenerate $J_i$ maximum $R$
times if needed

 \qquad\qquad\qquad   for $k=1:L_i$ \quad{//} go
through all elements in the $i$-th layer if conditions

\qquad\qquad\qquad\qquad\qquad\qquad\quad{//} (\ref{09}) are
satisfied

\qquad\qquad\qquad\qquad\qquad
 begin

\qquad\qquad\qquad\qquad\qquad\qquad \qquad // calculate energy on
$i$-th layer,

\qquad\qquad\qquad\qquad\qquad\qquad \qquad // calculate
polarization on $x,y$ and $z$-axis

\qquad\qquad\qquad\qquad\qquad\qquad \qquad // calculate $x_{i+1}$
and $y_{i+1},$

\qquad\qquad\qquad\qquad\qquad\qquad \qquad // save $J_i$ value

\qquad\qquad\qquad\qquad\qquad\qquad\qquad .\quad .\quad .\quad .

\qquad\qquad\qquad\qquad\qquad end

\qquad\qquad\qquad\quad endfor

\qquad\qquad endfor

\qquad  endfor

if ($i==N_x$) // reached the $N_x$-th layer

\qquad \qquad begin

\qquad \qquad \qquad    // save energy, polarizations values

\qquad \qquad end

endif\\
// construct distribution functions of energy $\varepsilon$, polarization $p$ and\\
// interaction constant $J$\\
\\
// calculate the mean value of energy $\bar{\varepsilon}$, polarization $\bar{p}$, interaction
constant $\bar{J}$ and \\
// its variance $\bar{J^2}$.

\section{Numerical Simulation}
We will consider an ideal ensemble of 1$D$ SSCs which consists of
$M$ number  of spin-chains each of them with the length 25$d_0$. For
realization of parallel simulation we will use  algorithm \textbf{A}
(see FIG 2).

The parallel algorithm works in the following way. Randomly $M$ sets
of initial parameters are generated and parallel calculations of
equations  (\ref{20}) for unknown variables $x$ and $y$ transact
with taking into account conditions (\ref{21}). However only
specifying of initial conditions is not enough for solution of these
equations. Evidently these equations can be   solved after
definition of the constant $J_{0\,1}$,  which is also randomly
generated.
 In the case when solutions are
found then conditions of stability of spin in node (\ref{09}) are
checked. The solution proceeds for the following spin if the
specified conditions (\ref{09})  are satisfied. If conditions are
not satisfied, a new constant $J_{0\,1}$ is randomly generated and
correspondingly new solutions are found which are checked later  on
conditions (\ref{09}). This cycle on each spin repeats until the
solutions do not satisfy  to conditions of the minimum spin energy
in the node.

\begin{figure}\center
\includegraphics[height=140mm,width=120mm]{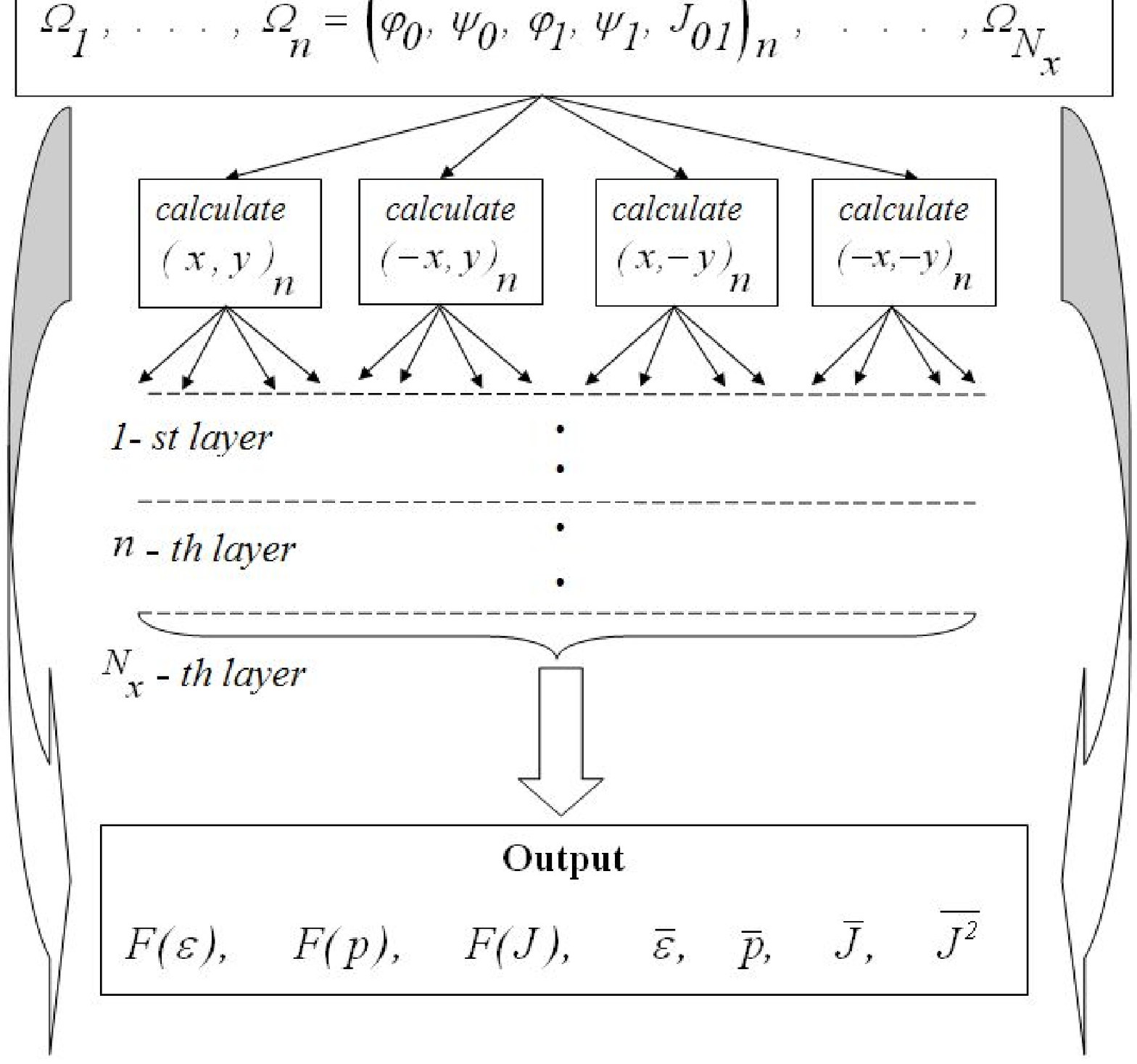}
\vspace{-5.5 cm}\caption{The algorithm of $1D$ SSCs of ideal
ensemble parallel simulation of statistical parameters.}
\label{fig2}
\end{figure}

At  first we have conducted numerical simulation for definition of
different statistical parameters of the ensemble which  consists of
$10^2$ spin-chains. Let us recall that the number of simulation of
spin-chains define the number of spin-chains in the ensemble. As the
simulation shows (see the left picture in FIG 3)  the energy
distribution function has a set of local maximums
($\varepsilon^{(0)},...,\varepsilon^{(m)})$. Obviously they are
dimensional effects  and  are similar to the first-order phase
transitions which often happen in spin-glass systems \cite{Bind}).

\begin{figure}
\center
\includegraphics[height=135mm,width=130mm]{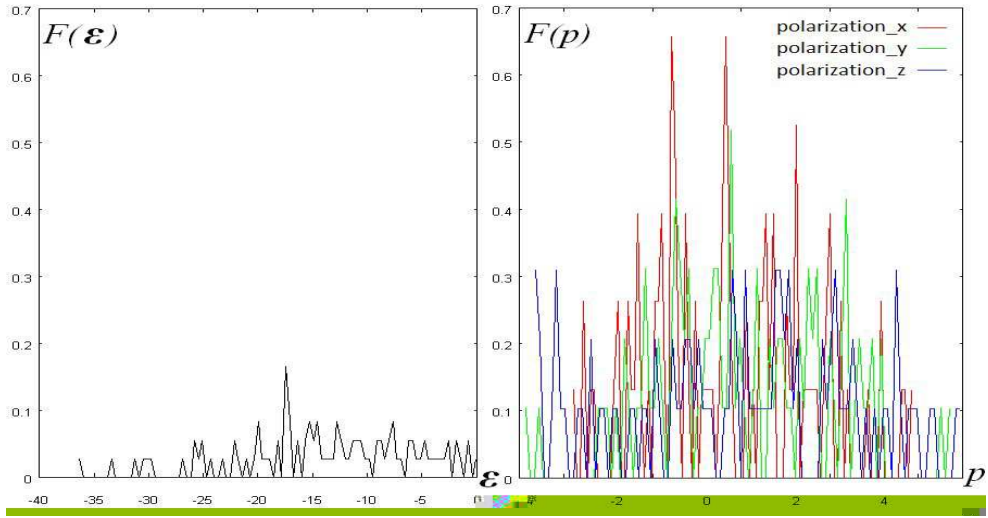}
 \vspace{-5.2 cm}\caption{The energy distribution where there are apparently  many local minimum
of energy for ensemble of 1$D$ SSCs  with the length of $L_x=25d_0$,
which consists of $10^2$ spin-chains (the left picture). On the
right picture  polarization distributions of ensemble on coordinates
$x,y$ and $z$ are shown.} \label{fig3}
\end{figure}

\begin{figure}
\center
\includegraphics[height=135mm,width=130mm]{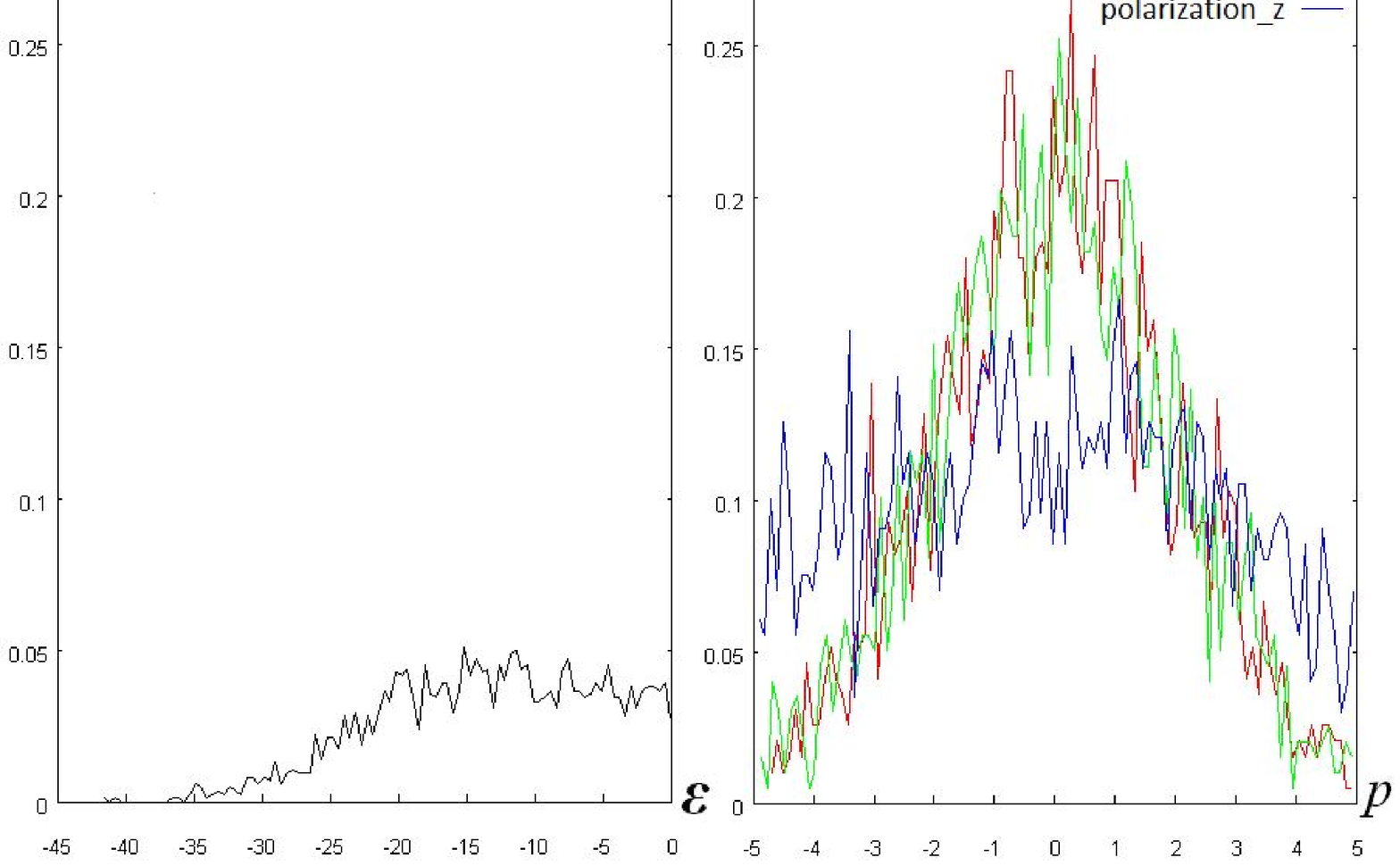}
 \vspace{-5.2 cm}\caption{In the left picture
is shown the energy distribution in the ensemble of 1$D$ SSCs with
the length of $L_x=25d_0$, which consists of 2$\cdot10^3$
spin-chains. Apparently,  the number of local minimums of energy is
promptly reduced comparing with the  increase of spin-chains. On the
right picture  polarization distributions of ensemble on coordinates
$x,y$ and $z$ are shown.} \label{fig4}
\end{figure}

Let us note that during simulation we suppose that spin-chains can
be polarized up to 20 percent i.e. the total value  of spins sum in
each chain  can be in an interval of $-5 \leq  p\leq 5$, where $p$
designates the polarization of spin-chain. In other words each
spin-chain is a vector of certain length which  is directed to
coordinate $x$.  As calculations show, in the ensemble consisting of
a small number of spin-chains, for example, of the order $10^2$, the
self-averaging of spin-chains  does not  occur  in full measure i.e.
the total polarization of an ensemble differs from zero:
$p_x=-0.33099,\, p_y=-0.035191,\, p_z=-0.024543$ where
$p=\int_{-\infty}^{+\infty} F(p)dp$, where it is supposed that
$p=(p_x,\,p_y,\,p_z)$. In this case the average energy of an
ensemble is equal to $\bar{\varepsilon}=-14.121$, where
$\bar{\varepsilon}=\int_{-\infty}^0 F(\varepsilon)\varepsilon
d\varepsilon$.

For the ensemble which consists of $2.10^3$ spin-chains (see FIG 4),
the dimensional effects practically disappear. The summary
polarization of ensemble in this case is very small: $p_x=-0.020538,
\,p_y=-0.047634,\, p_z=-0.12687$ and correspondingly the average
energy of $1D$ SSC is equal to $\bar{\varepsilon}=-13.603$.

Ensemble which consists of $10^4$ spin-chains has an energy
distribution $F(\varepsilon)$ with one global maximum (see Fig 5).
As to polarization distributions, $F(p_x)$ $F(p_y),$ and $F(p_z)$,
in the considered case are obviously very symmetric in comparison
with similar distributions of previous ensembles (see FIG 3 and Fig
4). The average values of polarizations on coordinates for this
ensemble are much smaller $p_x=-0.0072863,\, p_y=- 0.014242,\,
p_z=-0.018387$, correspondingly the average energy is equal to
$\bar{\varepsilon}=-13.634$. Thus in the case when ensemble consists
of a big number of spin-chains,  the self-averaging of spin-chains
system occurs with high accuracy. Whereas the summation procedure on
the number of spins in chain or spin-chains ensemble is similar to
the procedure of averaging by the natural parameter or "timing" in
the dynamical system, it is possible to introduce  the concept of
ergodicity for the both separate spin-chains and ensemble as a
whole.

\begin{figure}
\center
\includegraphics[height=135mm,width=130mm]{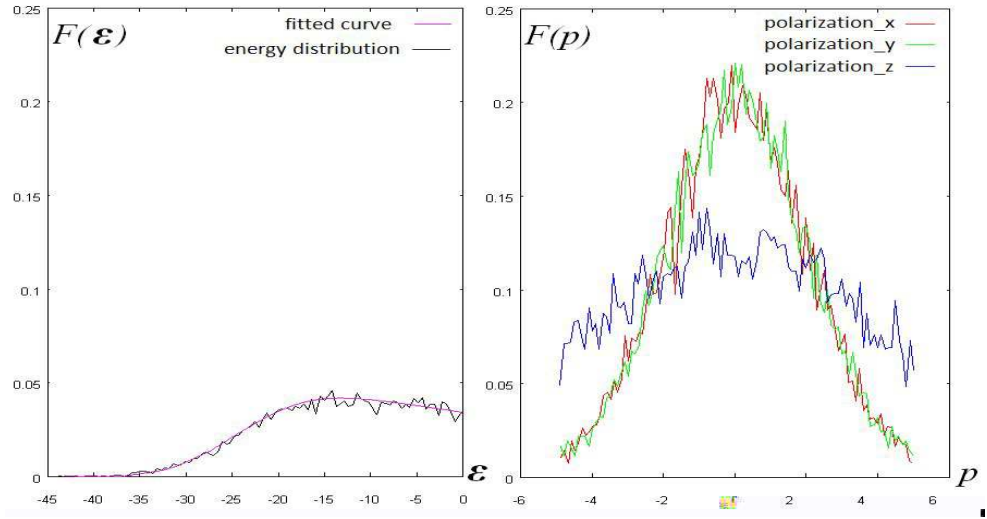}
 \vspace{-5.5 cm}\caption{The energy distribution and its fitted curve (left
picture) in ensemble of 1$D$ SSCs  with the length of $L_x=25d_0$,
which consists of $10^4$ spin-chains. Evidently there is only one
global maximum for energy distribution. In the right picture
polarization distributions are shown correspondingly on coordinates
$x,y$ and $z$.} \label{fig5}
\end{figure}
Thus as calculations show Birkhoff ergodic hypothesis
\cite{Birkhoff} may be used for ensembles which consist of $M\sim
N_x^2$ spin-chains in order to change the summation of spin-chains
on the  integration by the energy distribution of the ensemble. The
energy distribution of ensemble does not depend on the length of the
spin-chain in the limit of ergodicity and it can be fitted very
precisely with Eckart function \cite{Eckart} (see FIG 5, the smooth
\begin{equation}
F(\varepsilon)=C(a,b,c,\gamma)\biggl\{\frac{a}{b+e^{-2\gamma
\varepsilon}}+\frac{c\gamma^2}{(e^{-\gamma \varepsilon}+e^{\gamma
\varepsilon})^2}\biggr\},
 \label{22}
\end{equation}
where $a,b,c$ and $\gamma$ some constants, in addition $C$ is a
normalization constant  and  can be found from the condition:
\begin{equation}
 \int_{-\infty}^0F(\varepsilon)d\varepsilon=1.
\label{23}
\end{equation}
By placing (\ref{22}) into (\ref{23}) we can find:
\begin{equation}
C^{-1}(a,b,c,\gamma)=\frac{a}{2b\gamma}\ln(1+b)+\frac{c\gamma}{4}.
\label{24}
\end{equation}
After  fitting the energy distribution by means of analytical
function (\ref{22}) we find values of constants by entering into the
function: $a=131.4,\,b=3138.2,\,\,c=-1.20344$ and $\gamma=0.162174.$

\begin{figure}
\vspace{-0.81 cm}\center
\includegraphics[height=135mm,width=130mm]{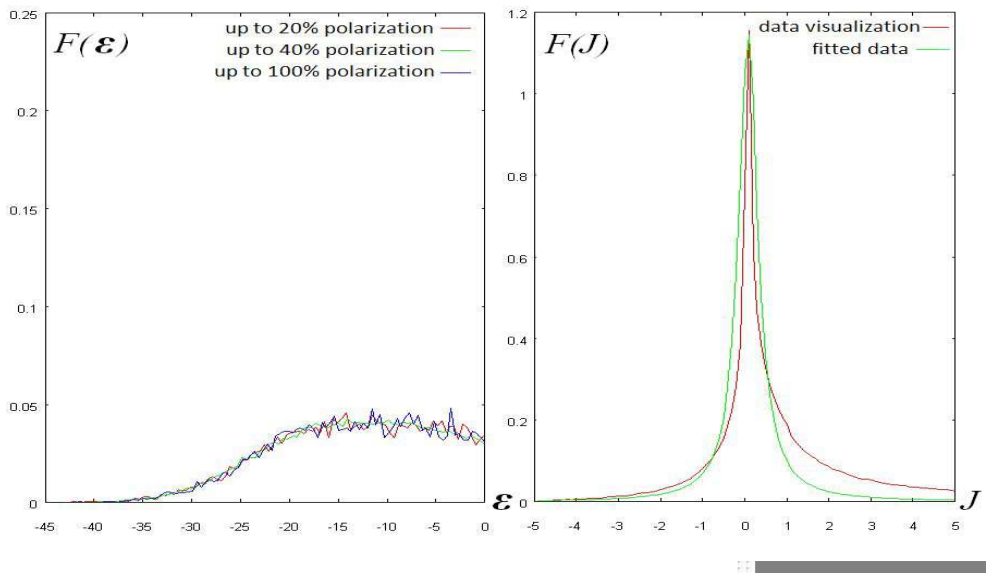}
 \vspace{-5.0 cm}\caption{The energy distributions  for ensembles  consisting of 1$D$
SSCs  of the length  $L_x=25d_0$,  with spin-chains polarization
correspondingly  up to $20,\,40$ and $100$ percents  (left picture).
Note that all the ensembles consist of $10^4$ spin-chains and
 their distributions practically do not differ. On the
right picture  the distribution of the spin-spin interaction
constant is shown which differs essentially   from
Gauss-Edwards-Anderson distribution model (2). } \label{fig6}
\end{figure}

We have also calculated 1$D$ SSCs ensemble with the length of
spin-chains 25$d_0$ and correspondingly with polarizations of
spin-chains  up to 20,\, 40,\, and 100 percents (see Fig 6, the left
picture). In particular, as it follows from the picture the energy
distribution does not depend on the degree of spin-chains
polarization. We also have conducted simulation of ensembles which
consist of spin-chains with lengths $100d_0$ and $1000d_0$
correspondingly. As the numerical modeling shows, statistical
properties of ensembles are similar.  In the considered cases
distributions of energy concentrate correspondingly  on scales
$100d_0$ and $1000d_0$. Limits of ergodicities of  ensembles are
also investigated  and it is shown that in these cases too it is of
an order $N_x^2$.

Finally  it is important to note that the distribution of spin-spin
interaction constant is not  defined apriori with the help of
expression (\ref{02}) but  with the mass calculations of equations
(\ref{06}). On the basis of the obtained numerical data, the
distribution of interaction constant $W(J)\equiv F(J)$ is
constructed (see Fig 6, the right picture) from which it follows,
that it essentially differs from the Gauss-Edwards-Anderson
distribution model (\ref{02}). The obtained distribution relatively
is   well fitted by the normalized to the unit of nonsymmetric
Cauchy function \cite{Spiegle}:
\begin{equation}
F(J)=\frac{g+\beta J}{\pi \bigl[g^2+(J-a_0)^2\bigr]}.
 \label{25}
\end{equation}
where $g,\,\beta$ and $a_0$ are some  adjusting parameters which are
found from the condition of a good approximation of the data
visualization curve. In the considered case they are correspondingly
equal to: $g=0.27862,\,\beta=0.009$ and $a_0=0.083236$.
Nevertheless, as the detailed analysis of curve of numerical data
visualization shows (in particular its asymptotes)  the distribution
of interaction constant can be approximated precisely by Le\'{v}y
skew alpha-stable distribution function.  Let us recall that
Le\'{v}y skew alpha-stable distribution is a continuous probability
and a limit of certain random process
$X(\alpha,\beta,\gamma,\delta;k)$ where parameters describe
correspondingly: an index of stability or characteristic exponent $
\alpha\in (0;2]$, a skewness parameter $\beta\in [-1;1]$, a scale
parameter $\gamma > 0$, a location parameter $\delta \in \mathbb{R}$
and an integer $k$  shows the certain parametrization (see in more
detailed references \cite{Ibragimov,Nolan}).  Let us note, that the
mean of distribution and its  variance are infinite. However, taking
into account that spin-spin interaction constant has limited value
in real physical systems, it is possible to calculate distribution
mean and its variance. In particular if $J\in [-5,+5]$ then
$\overline{J}=0.50113$ and $\overline{J^2}=2,1052$.

\section{Conclusion}
The investigation of statistical properties  of classical spin-glass
system of various sizes is very important for understanding
possibilities of effective influence and control  over parameters of
medium with the help of weak external fields.  Evidently, when we
put the spin-glass in external field the space-time periods define
scales on which probably an essential changes in medium occur. For
simplicity we suppose that the spin-glass system  is an ensemble
which consists of disordered 1$D$ steric spin-chains of $L_x$
lengths, between which interaction is absent (ideal ensemble). This
type of classical ensemble is described by  Heisenberg Hamiltonian
(\ref{02}). We have researched conditions of arising of stable
spin-chains Eqs. (\ref{11}) and nonequalities (\ref{09}) and found a
latent  connection between random variables (see expression
(\ref{19})), which shows that the distribution for spin-spin
interaction constant can not be described by Guss-Edwards-Anderson
model. In  the result of equations of stationary points analysis
(\ref{11}) we have found system of recurrent equations (\ref{20})
and new conditions (\ref{21}). On the basis of obtained mathematical
formulas the effective parallel algorithm  for numerical simulation
is developed which was realized on the example of the ensemble which
consists  of 1$D$ SSCs with length 25$d_0$. Similar to the dynamical
systems, we have introduced the idea of Birkhoff ergodic hypothesis
\cite{Birkhoff} for the statical spin-glass systems. In this case
the number of spin-chains of ensemble plays a role of the natural or
"timing" parameter of the system. Numerical simulations show that
the ergodic hypothesis may be used for the case when ensemble
consists of $M\propto N_x^2$ spin-chains in order to change the
summation of spin-chains  on the integration by the energy
(polarization, etc.) distribution of the ensemble.

In particular, we have made numerical experiments for ensembles
which include  $10^2$,\,\, $2\cdot 10^3$ and $10^4$ spin-chains.  As
it was shown by simulations in the case when $M\ll N_x^2$ for an
ensemble, they are characteristic dimensional effects in energy
distribution (the left picture on FIG 3). When the number of
spin-chains is $M$ of order $2\cdot 10^3$ or more $10^4$,
dimensional effects disappear and correspondingly energy
distribution functions have one global maximum (see left pictures on
FIG 4 and FIG 5). As it was shown, when increasing spin-chains
number, the total and partial polarizations of the ensemble
disappear. Let us note, that at modelling by algorithm (see  scheme
on FIG 2) condition (\ref{19}) specifies the region of localization
of random interaction constant  $J_{i\,i+1}$ which depends on
angular configurations $(i-1)$-th and $i$-th spins and interaction
constant $J_{i-1\,i}$ between them.  As a result, it allows to
accelerate calculations of each spin-chain and hence the speed of
parallel calculations of ensemble is increased essentially.

Finally it is important to note that it is proved, that the
spin-spin interaction constant $J_{i\,i+1}$ has a form of Le\'{v}y
skew alpha-stable distribution (see the right picture on FIG 6). The
considered scheme of  solution of 1$D$ steric spin-glass problem can
be used in different applied fields (see e.g. \cite{Helmut}). It can
also be useful for analyzing 3$D$ spin-glass problem and creation of
an effective parallel simulation algorithm of the spin-glass system
with large dimensionality.


\begin{thebibliography}{099}
\bibitem{Bind}  K. Binder  and  A. P. Young, Spin glasses:
Experimental facts, theoretical concepts, and open questions. Rev.
Mod. Physics, {\bf{58}}(4),  801-976 (1986).

\bibitem{Mezard} M. M\'{e}zard, G. Parisi, M. A. Virasoro, Spin Glass Theory and Beyond (World
Scientific, Singapore, 1987)

\bibitem{Young} A. P. Young (ed.), Spin Glasses and
Random Fields (World Scientific, Singapore, 1998)

\bibitem{Fisch} R. Fisch and A. B. Harris, Spin-glass model in
continuous dimensionality, Phys. Rev. Let., \textbf{47}, 620 (1981).

\bibitem{Bov} A. Bovier, Statistical Mechanics of Disordered Systems:
A Mathematical Perspective, Cambridge Series in Statistical and
Probabilistic Mathematics, p 308 (2006).

\bibitem{Tu}
Y. Tu, J. Tersoff and G. Grinstein, Structure and Energetic of the
$Si$ and $SiO_2$ Interface, Phys. Rev. Lett., \textbf{81}, 4899
(1998).

\bibitem{Chary} K. V. R. Chary, G. Govil, NMR in Biological Systems:
From Molecules to Human (Focus on Structural Biology 6), Springer, p
511, (2008).


\bibitem{Baake} E. Baake, M. Baake and H. Wagner, Ising Quantum Chain
is a Equivalent to a Model of Biological Evolution, Phys. Rev. Let.,
\textbf{78}(3), 559-562 (1997.)

\bibitem{gev} A S Gevorkyan  et al.,
New Mathematical Conception and Computation Algorithm for Study of
Quantum 3D Disordered Spin System Under  the Influence of External
Field, Trans. On Comput. Sci., VII, LNCS 132-153, Spinger-Verlage,
10.1007/978-3-642-11389-58

\bibitem{EdwAnd} S. F. Edwards and P. W. Anderson, Theory of spin
glasses, J. Phys. F {\textbf{9}}, 965 (1975).



\bibitem{Birkhoff} J. von Neuman, Physical Applications of the
Ergodic Hypothesis, Proc. Nat. Acad. Sci. USA, \textbf{18}(3):
263-266 (1932).\\  G. D. Birkhoff, What is ergodic theorem? American
Mathematical Monthly, \textbf{49}(4): 222-226 (1931).

\bibitem{Eckart} S. Fl\"{u}gge, Practical Quantum Mechanics I,
(Springer-Verlag, Berlin-Heidelberg- New York 1971).

\bibitem{Spiegle} M. R. Spiegle, Theory and Problems of Probability
and Stochastics, (New-York, McGraw-Hill, pp 114-115, 1992).

\bibitem{Ibragimov} I. Ibragimov and Yu. Linnik, Independent and
Stationary Sequences of Random Variebles, (Wolters-Noordhoff
Publishing Groningen, The Netherlands 1971).

\bibitem{Nolan} J. P. Nolan, Stable Distributions: Models for Heavy
Tailed Data  (2009-02-21). $en.wikipedia.org/Stable_/distribution$.

\bibitem{Helmut} H. G. Katzgraber, A. K. Hartmann and A. P. Young,
New Insights from One-Dimensional Spin Glasses, (2008)
ArXiv:0803.3417v1 [cond-mat.dis-nn].
\end{thebibliography}
\end{document}